\documentclass[preprint]{aastex}

\slugcomment{Draft v5.3, October 27}

\shorttitle{Pulsed Gamma Rays from PSRs B1937+21 and B1957+20}
\shortauthors{Guillemot et al.}

\begin{document}

%\linenumbers


\title{Pulsed Gamma Rays from the Original Millisecond and Black Widow Pulsars: a case for Caustic Radio Emission?}

\author{
L.~Guillemot\altaffilmark{1,2},
T.~J.~Johnson\altaffilmark{3,4,5,6},
C.~Venter\altaffilmark{7,8},
M.~Kerr\altaffilmark{9,10},
B.~Pancrazi\altaffilmark{11,12},
M.~Livingstone\altaffilmark{13},
G.~H.~Janssen\altaffilmark{14},
P.~Jaroenjittichai\altaffilmark{14},
M.~Kramer\altaffilmark{14,1},
I.~Cognard\altaffilmark{15},
B.~W.~Stappers\altaffilmark{14},
A.~K.~Harding\altaffilmark{3},
F.~Camilo\altaffilmark{16},
C.~M.~Espinoza\altaffilmark{14},
P.~C.~C.~Freire\altaffilmark{1},
F.~Gargano\altaffilmark{17},
J.~E.~Grove\altaffilmark{18},
S.~Johnston\altaffilmark{19},
P.~F.~Michelson\altaffilmark{9},
A.~Noutsos\altaffilmark{1},
D.~Parent\altaffilmark{20},
S.~M.~Ransom\altaffilmark{21},
P.~S.~Ray\altaffilmark{18},
R.~Shannon\altaffilmark{19},
D.~A.~Smith\altaffilmark{22},
G.~Theureau\altaffilmark{15},
S.~E.~Thorsett\altaffilmark{23},
N.~Webb\altaffilmark{11,12}
}
\altaffiltext{1}{Max-Planck-Institut f\"ur Radioastronomie, Auf dem H\"ugel 69, 53121 Bonn, Germany}
\altaffiltext{2}{email: guillemo@mpifr-bonn.mpg.de}
\altaffiltext{3}{NASA Goddard Space Flight Center, Greenbelt, MD 20771, USA}
\altaffiltext{4}{Department of Physics and Department of Astronomy, University of Maryland, College Park, MD 20742, USA}
\altaffiltext{5}{National Research Council Research Associate, National Academy of Sciences, Washington, DC 20001, resident at Naval Research Laboratory, Washington, DC 20375, USA}
\altaffiltext{6}{email: tyrel.j.johnson@gmail.com}
\altaffiltext{7}{Centre for Space Research, North-West University, Potchefstroom Campus, Private Bag X6001, 2520 Potchefstroom, South Africa}
\altaffiltext{8}{email: Christo.Venter@nwu.ac.za}
\altaffiltext{9}{W. W. Hansen Experimental Physics Laboratory, Kavli Institute for Particle Astrophysics and Cosmology, Department of Physics and SLAC National Accelerator Laboratory, Stanford University, Stanford, CA 94305, USA}
\altaffiltext{10}{email: kerrm@stanford.edu}
\altaffiltext{11}{CNRS, IRAP, F-31028 Toulouse cedex 4, France}
\altaffiltext{12}{GAHEC, Universit\'e de Toulouse, UPS-OMP, IRAP, Toulouse, France}
\altaffiltext{13}{Department of Physics, McGill University, Montreal, PQ, Canada H3A 2T8}
\altaffiltext{14}{Jodrell Bank Centre for Astrophysics, School of Physics and Astronomy, The University of Manchester, M13 9PL, UK}
\altaffiltext{15}{ Laboratoire de Physique et Chimie de l'Environnement, LPCE UMR 6115 CNRS, F-45071 Orl\'eans Cedex 02, and Station de radioastronomie de Nan\c{c}ay, Observatoire de Paris, CNRS/INSU, F-18330 Nan\c{c}ay, France}
\altaffiltext{16}{Columbia Astrophysics Laboratory, Columbia University, New York, NY 10027, USA}
\altaffiltext{17}{Istituto Nazionale di Fisica Nucleare, Sezione di Bari, 70126 Bari, Italy}
\altaffiltext{18}{Space Science Division, Naval Research Laboratory, Washington, DC 20375-5352, USA}
\altaffiltext{19}{CSIRO Astronomy and Space Science, Australia Telescope National Facility, Epping NSW 1710, Australia}
\altaffiltext{20}{Center for Earth Observing and Space Research, College of Science, George Mason University, Fairfax, VA 22030, resident at Naval Research Laboratory, Washington, DC 20375, USA}
\altaffiltext{21}{National Radio Astronomy Observatory (NRAO), Charlottesville, VA 22903, USA}
\altaffiltext{22}{Universit\'e Bordeaux 1, CNRS/IN2p3, Centre d'\'Etudes Nucl\'eaires de Bordeaux Gradignan, 33175 Gradignan, France}
\altaffiltext{23}{Department of Physics, Willamette University, Salem, OR 97031, USA}


\begin{abstract}

We report the detection of pulsed gamma-ray emission from the fast millisecond pulsars (MSPs) B1937+21 (also known as J1939+2134) and B1957+20 (J1959+2048) using 18 months of survey data recorded by the \emph{Fermi} Large Area Telescope (LAT) and timing solutions based on radio observations conducted at the Westerbork and Nan\c cay radio telescopes. In addition, we analyzed archival \emph{RXTE} and \emph{XMM-Newton} X-ray data for the two MSPs, confirming the X-ray emission properties of PSR B1937+21 and finding evidence ($\sim 4\sigma$) for pulsed emission from PSR~B1957+20 for the first time. In both cases the gamma-ray emission profile is characterized by two peaks separated by half a rotation and are in close alignment with components observed in radio and X-rays. These two pulsars join PSRs J0034$-$0534 and J2214+3000 to form an emerging class of gamma-ray MSPs with phase-aligned peaks in different energy bands. The modeling of the radio and gamma-ray emission profiles suggests co-located emission regions in the outer magnetosphere.

\end{abstract}


\keywords{gamma rays: observation -- pulsars: general -- pulsars: individual (PSR B1937+21, PSR B1957+20) -- radiation mechanisms: non-thermal}

\section{Introduction}

The Large Area Telescope (LAT) aboard the \emph{Fermi Gamma-ray Space Telescope} has firmly established millisecond pulsars (MSPs), rapidly-rotating neutron stars ($P \lesssim$ 30 ms) with small rotational spin-downs ($\dot P \lesssim 10^{-17}$), as sources of GeV gamma rays. Nine MSPs known prior to the \emph{Fermi} mission have so far been observed to emit pulsed gamma rays \citep{FermiJ0030, Fermi8MSPs, FermiJ0034}, and radio searches at the position of \emph{Fermi} LAT unassociated sources, such as those in the \emph{Fermi} LAT First Year Catalog \citep{Fermi1FGL}, have led to the discovery of over thirty previously unknown MSPs \citep[see e.g.][]{Hessels2011}. Six of these new pulsars have been shown to emit pulsed gamma rays already \citep{Cognard2011,Keith2011,Ransom2011}. The LAT has also detected gamma-ray emission from several globular clusters, and the observed properties are consistent with the summed contribution of a population of MSPs \citep{Fermi47Tuc,Kong2010,FermiGCs}. In addition, the \emph{AGILE} telescope reported a 4.2$\sigma$ detection of PSR B1821$-$24 in the globular cluster M28 in gamma rays \citep{Pellizzoni2009b}. These different observations indicate that MSPs are prominent sources of gamma rays and that many of them are awaiting detection with the \emph{Fermi} LAT.

All MSPs detected by \emph{Fermi} to date are relatively energetic, with spin-down luminosities $\dot E = 4 \pi^2 I \dot P / P^3 > 10^{33}$ erg s$^{-1}$ (where $I$ denotes the moment of inertia, assumed to be $10^{45}$ g cm$^2$ in this work), making PSRs B1937+21 ($\dot E = 1.1 \times 10^{36}$ erg s$^{-1}$) and B1957+20 ($\dot E = 7.5 \times 10^{34}$ erg s$^{-1}$) good candidates for detection in gamma rays with \emph{Fermi}. Nevertheless, the two MSPs are more distant than the bulk of gamma-ray-detected MSPs \citep[see][]{FermiPSRCatalog}, and are located at low Galactic latitudes and therefore suffer from strong contamination from the diffuse Galactic emission (the properties of these two MSPs are listed in Table \ref{proprietes}), making these pulsars difficult to detect. 

In this article we describe the detection of pulsed gamma-ray emission from PSRs B1937+21 and B1957+20 using the first 18 months of data recorded by the \emph{Fermi} LAT. In addition, we analyzed the X-ray properties of the two MSPs using archival \emph{RXTE} and \emph{XMM-Newton} data. The two MSPs are observed to emit radio, X-rays, and gamma rays, in near alignment. We present results of the modeling of these radio and gamma-ray components in the context of geometrical models of emission from pulsar magnetospheres. Additionally, we examine the possibility that gamma rays are produced by colliding winds in the PSR~B1957+20 system, as was observed in X-rays \citep{Stappers2003}. 

\section{PSRs B1937+21 and B1957+20}

PSR~B1937+21 was the first MSP ever discovered \citep{BackerJ1939}, and remained the pulsar with the shortest known rotational period ($P \sim 1.558$ ms) until the recent detection of a 1.396 ms pulsar in the globular cluster Terzan 5, PSR~J1748$-$2446ad \citep{HesselsJ1748ad}. With a pulse period of 1.607 ms, the first ever ``black widow'' pulsar discovered, PSR~B1957+20 \citep{Fruchter1988}, has the third shortest rotational period of currently known pulsars. The distances to these MSPs are relatively uncertain: the NE2001 model of Galactic electron density \citep{NE2001} places PSR~B1937+21 at $d = 3.6 \pm 1.4$ kpc, assuming a 40\% uncertainty in the dispersion model \citep{Brisken2002}. However, timing measurements by \citet{Verbiest2009} led to $\pi = 0.13 \pm 0.07$ mas, and therefore $d = 7.7 \pm 3.8$ kpc. The only distance estimates currently available for PSR~B1957+20 are based on the dispersion measure (DM) and the models of Galactic free electron density. The NE2001 model places this MSP at $d = 2.5 \pm 1.0$ kpc. This estimate is supported by spectral analyses of the binary companion at optical wavelengths, placing a lower limit on the distance of $d \gtrsim 2$ kpc \citep{vanKerkwijk2011}. In the following, we will use the parallax distance from \citet{Verbiest2009} of $7.7$ kpc for PSR~B1937+21 and the NE2001 distance of $2.5$ kpc for PSR~B1957+20. 

The proper motion of PSR~B1937+21 is also relatively uncertain: available estimates based on pulsar timing measurements range from $\mu_T = \sqrt{ \mu_\alpha^2 \cos^2 \delta + \mu_\delta^2 } = 0.28 \pm 0.09$ mas yr$^{-1}$ \citep{Verbiest2009phd} to $\mu_T = 1.6 \pm 0.2$ mas yr$^{-1}$ \citep{Hotan2006}, while \citet{Kaspi1994}, \citet{Cognard1995}, and \citet{Verbiest2009} give intermediate values with comparably small uncertainties, possibly underestimated. However, even with the largest of the measurements, the period derivative is weakly affected by the kinematic Shklovskii effect, which makes the apparent $\dot P$ greater than the intrinsic value by $(P \mu_T^2 d) / c$ \citep{Shklovskii1970}. At a distance of 7.7 kpc and assuming the largest proper motion value of $1.6$ mas yr$^{-1}$, the expected Shklovskii contribution is $7 \times 10^{-23}$, negligible compared to the apparent $\dot P$ of $\sim 1.05 \times 10^{-19}$. In this paper we will use the intermediate proper motion value measured by \citet{Cognard1995} of $0.80 \pm 0.02$ mas yr$^{-1}$. In contrast with PSR~B1937+21, the apparent period derivative of PSR~B1957+20 is strongly affected by the Shklovskii effect: with its distance of 2.5 kpc and transverse velocity of $\mu_T = 30.4 \pm 0.6$ mas yr$^{-1}$ \citep{Arzoumanian1994}, the Shklovskii effect decreases the $\dot P$ value by more than half. Table \ref{proprietes} lists characteristic properties of PSRs B1937+21 and B1957+20. The spin-down power $\dot E$ and magnetic field at the light cylinder $B_{LC}$ values are the highest among known Galactic MSPs. 

These two pulsars have been extensively studied at high energies: in X-rays, PSR~B1937+21 has a two-peaked profile similar to its radio profile, with peaks in close alignment with the giant radio pulse emission regions, lagging the normal radio emission peaks slightly \citep{Cusumano2003}. The X-ray emission from this pulsar is non-thermal, which distinguishes it from many other MSPs. The high $\dot E$ of PSR~B1957+20 would seemingly make this pulsar a good candidate for detection of X-ray pulsations. However, the X-ray emission from the system clearly has a strong contribution from the interaction of the pulsar wind with the companion, which is modulated at the orbital period \citep{Stappers2003}. Previous searches for X-ray pulsations \citep[see e.g.][]{Huang2007} have been unsuccessful.

\section{Observations and data analysis}

\subsection{Radio timing observations}
\label{ephemerides}

The timing solutions predicting rotational and orbital behaviors of PSRs B1937+21 and B1957+20 as a function of time were obtained from radio timing measurements contemporaneous with the first 18 months of the \emph{Fermi} mission. With spin-down energies above 10$^{34}$ erg s$^{-1}$, the two MSPs have been monitored by radio telescopes around the world as part of the pulsar timing campaign for \emph{Fermi} \citep{FermiTiming}. For PSR~B1937+21, we have built a timing solution by using 80 Times of Arrival (TOAs) taken at the Nan\c cay radio telescope in France \citep{Cognard2011} at 1.4 GHz between 2008 May 27 and 2010 February 7 and with a mean uncertainty on the determination of individual TOAs of 45 ns, and 42 TOAs recorded at the Westerbork Synthesis Radio Telescope (WSRT) \citep{Voute2002,Karuppusamy2008} at 1.4 and 2.3 GHz between 2008 January 27 and 2009 November 25, with a mean uncertainty of 579 ns. For~PSR B1957+20, the timing solution was built using 38 TOAs recorded between 2006 May 14 and 2009 December 19 at the Nan\c cay radio telescope at 1.4 GHz with a mean uncertainty of 2.3 $\mu$s, and 426 WSRT TOAs recorded at 0.35 GHz between 2008 January 27 and 2010 January 30 with a mean uncertainty of 2.8 $\mu$s.  

Ephemerides were built using the \textsc{Tempo2} pulsar timing package\footnote{http://tempo2.sourceforge.net/} \citep{tempo2}. The resulting ephemeris for PSR~B1937+21 gives a Root Mean Square (RMS) of the timing residuals of 197 ns. The Dispersion Measure (DM), necessary for the relative phasing of observations at different wavelengths, was fitted using the multi-frequency radio TOAs in the case of PSR~B1937+21. We measured a DM value for this pulsar of $71.01931 \pm 0.00019$ pc cm$^{-3}$ across the observation, where the error bar is the nominal 1$\sigma$ uncertainty reported by \textsc{Tempo2}. \citet{Cognard1995} measured a DM time derivative of $-0.0012 \pm 0.0001$ pc cm$^{-3}$ yr$^{-1}$. In our analysis we found no indication of significant time variation. In the case of PSR~B1957+20, however, significant DM variations with time were observed and needed to be taken into account when building the timing solution. The dispersion measure for PSR~B1957+20 was measured by generating one WSRT TOA per band of 20 MHz for each observation, thereby producing 8 TOAs per observation. TOAs corresponding to weak detections or where the pulsar was in eclipse were discarded. Assuming that for a given epoch pulses in the different frequency sub-bands were emitted simultaneously, we fitted the DM by measuring the time drift of TOAs as a function of radio frequency.  DM excursions of 0.0008 and $-$0.0011 around an average value of 29.1259 pc cm$^{-3}$ were observed. A DM of $29.12644 \pm 0.00029$ pc cm$^{-3}$ was measured at the epoch TZRMJD of $\sim$54550 MJD, the TZRMJD parameter defining a reference epoch at which the rotational phase predicted by the timing solution is 0.\footnote{See http://www.atnf.csiro.au/research/pulsar/ppta/tempo2/manual.pdf for more details}. DM values for each TOA in our dataset were determined by interpolating the values measured with the WSRT data. The TOAs and their corresponding DM values were then analyzed using \textsc{Tempo2}, resulting in an RMS of timing residuals of 4.1 $\mu$s. The latter RMS value is larger than the timing accuracy of the LAT of less than 1 $\mu$s \citep{FermiCalibration}, but negligible for the low-statistics gamma-ray light curves of PSR~B1957+20 (see Section \ref{gamma}). Similarly, the uncertainties on measured DM values of PSRs B1937+21 and B1957+20 correspond to errors in the extrapolation of 1.4 GHz TOAs to infinite frequency of 400 and 600 ns respectively, again negligible considering the low statistics. The timing solutions will be made available through the \emph{Fermi} Science Support Center\footnote{ http://fermi.gsfc.nasa.gov/ssc/data/access/lat/ephems/}. 

\subsection{Gamma-ray analysis}
\label{gamma}

\subsubsection{Initial searches for pulsations and spectral analysis}
\label{analyse_spectrale}

To search for gamma-ray pulsations from the two MSPs, we selected events recorded between 2008 August 4 and 2010 February 13, with energies above 0.1 GeV, zenith angles $\leq$ 105$^\circ$ and belonging to the ``Diffuse'' class of events under the P6\_V3 instrument response functions (IRFs), those events having the highest probabilities of being photons \citep{FermiLAT}. We excluded times when the rocking angle of the instrument exceeded 52$^\circ$, required that the DATA\_QUAL and LAT\_CONFIG are equal to 1 and that the Earth's limb did not infringe upon the Region of Interest (ROI). The gamma-ray data were analyzed using the \emph{Fermi} science tools (STs) v9r18p6\footnote{http://fermi.gsfc.nasa.gov/ssc/data/analysis/scitools/overview.html}. Photon dates were phase-folded using the \emph{Fermi} plug-in distributed with the \textsc{Tempo2} pulsar timing package\footnote{See http://fermi.gsfc.nasa.gov/ssc/data/analysis/user/Fermi\_plug\_doc.pdf} \citep{Ray2011} and the ephemerides described in Section \ref{ephemerides}. For ROIs of 0.8$^\circ$ radii around PSRs B1937+21 and B1957+20, we obtained values of the bin-independent \emph{H}-test parameter \citep{deJager2010} of 30.8 and 41.1 respectively, corresponding to pulsation significances of 4.6 and 5.4$\sigma$. These relatively low significances, for PSR B1937+21 in particular, prompted us to verify the pulsed nature of the observed gamma-ray signal with an alternative pulsation search technique. 

As discussed in \citet{Kerr2011weights}, the LAT sensitivity to gamma-ray pulsars can be improved by weighting each photon by its probability of originating from the considered pulsar, and by taking these weights into account in the calculation of the pulsation significance. Among other advantages, this method is efficient at discriminating background events, which is particularly important for these two MSPs located at low Galactic latitudes and therefore observed in the presence of intense background contamination. In order to calculate the photon probabilities, we analyzed the spectral properties of putative gamma-ray sources located at the positions of the two MSPs. The spectral analysis was done using a binned maximum likelihood method \citep{Mattox1996} as implemented in the \emph{pyLikelihood} python module in the \emph{Fermi} STs.  All point sources from the 1FGL catalog \citep{Fermi1FGL} within 15$^\circ$ of the radio location of PSR~B1937+21 were included in the model, as well as additional sources from a list internal to the LAT team, based on 18 months of data. All point sources were modeled with power-law spectra except for the two known gamma-ray pulsars in the ROI, PSRs J1954+2836 and J1958+2846 \citep{FermiBlindSearch,Parkinson2010}, which were modeled as exponentially cut off power laws, of the form:

\begin{eqnarray}
\frac{dN}{dE} = N_0 \left( \frac{E}{\mathrm{1 GeV}} \right)^{-\Gamma} \exp\left[ - \left( \frac{E}{E_c}\right)^\beta \right].
\label{spectralfunc}
\end{eqnarray}

In Equation (\ref{spectralfunc}), $N_0$ denotes a normalization factor, $\Gamma$ is the photon index, $E_c$ is the cutoff energy of the pulsar spectrum, while $\beta$ is a parameter determining the steepness of the exponential cutoff. Spectra measured so far by the \emph{Fermi} LAT for gamma-ray pulsars are generally well-described by simple exponential models, $\beta \equiv 1$. The Galactic diffuse emission was modeled using the \emph{gll\_iem\_v02} mapcube.  The extragalactic diffuse background and residual instrument background were modeled jointly using the \emph{isotropic\_iem\_v02} template\footnote{Both diffuse models are available through the \emph{Fermi Science Support Center} (FSSC) (see http://fermi.gsfc.nasa.gov/ssc/}. The normalizations and indices for all point sources within 8$^{\circ}$ of PSR B1937+21 were left free in the fit as well as cutoff energies $E_c$ for pulsars.  The normalizations of the diffuse components were also left free. In a first attempt at analyzing the spectra of the two MSPs we considered all photons with energies above 0.1 GeV. However, PSR~B1937+21 could not be detected with sufficient significance below 0.5 GeV, and for both pulsars the best-fit spectral parameters were strongly affected by the background emission. We therefore subsequently rejected the events with energies below 0.5 GeV. After a first iteration of the analysis, the 1FGL sources J1938.2+2125c and J1959.6+2047, located 21.7' and 0.5' away from PSRs B1937+21 and B1957+20 respectively, were no longer found to be significant gamma-ray sources. We thus removed them from the spectral model and made another iteration. This led us to the gamma-ray spectral parameters listed in Table \ref{gammaparam}, where the first errors quoted are statistical and the second ones are systematic, and were calculated by following the same procedure as above, but using bracketing IRFs for which the effective area has been perturbed by $\pm$10\% at 0.1 GeV, $\pm$5\% near 0.5 GeV, and $\pm$20\% at 10 GeV with linear extrapolations in log space between. Table \ref{gammaparam} also lists integrated photon fluxes $F$ and energy fluxes $G$ above 0.5 GeV. To enable comparison with previously observed gamma-ray MSPs we quote photon and energy fluxes extrapolated to lower energies, obtained by integrating Equation (\ref{spectralfunc}) for energies above 0.1 GeV. 

\begin{figure}[ht]
\begin{center}
\includegraphics[scale=0.68]{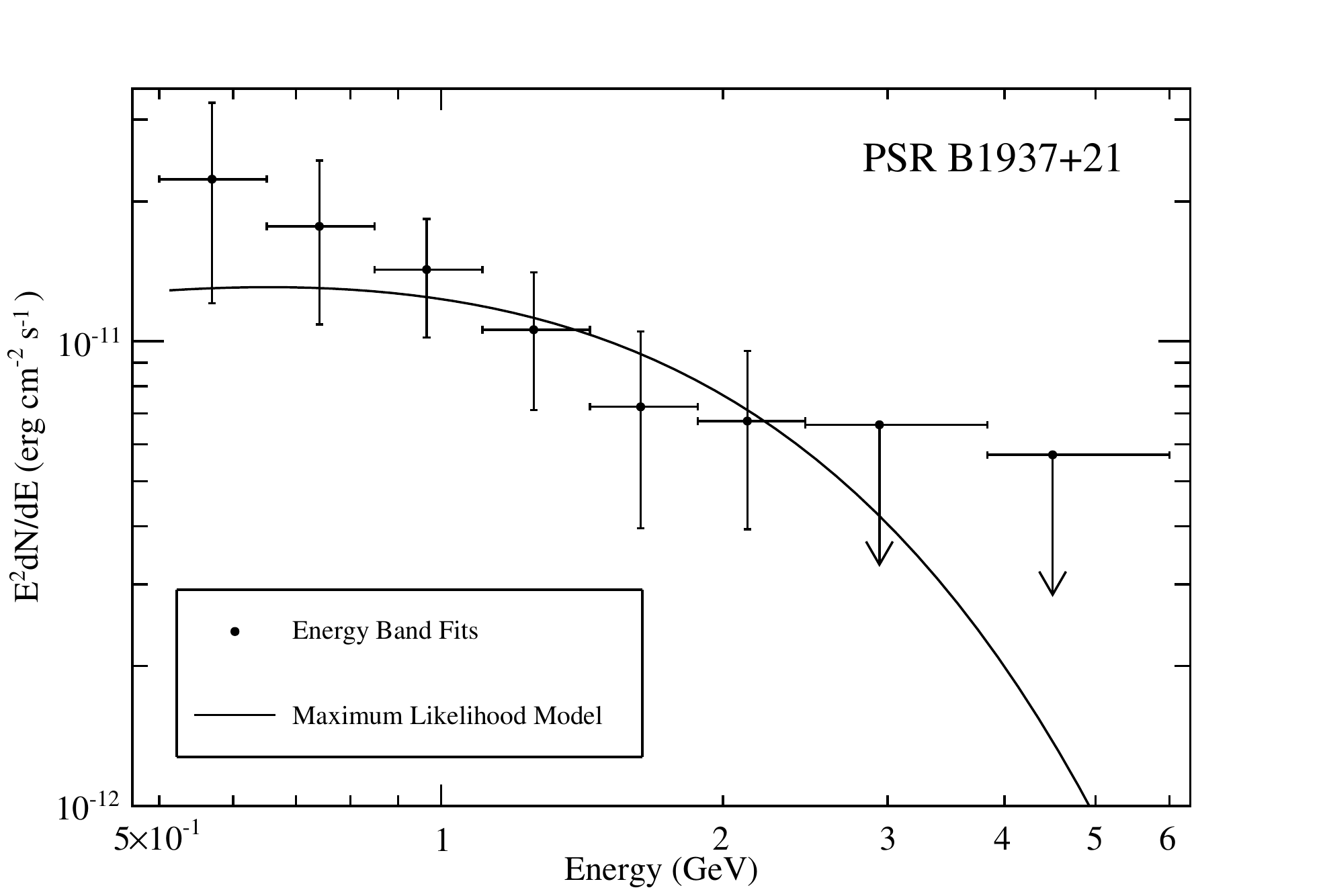}
\caption{Phase-averaged gamma-ray energy spectrum for PSR~B1937+21, above 0.5 GeV. The black line shows the best-fit model from fitting all the gamma-ray data above 0.5 GeV with the simple exponentially cutoff power-law functional form given in Equation (\ref{spectralfunc}). Data points are derived from likelihood fits of individual energy bands where the pulsar is modeled with a simple power-law form. A 95\% confidence level upper limit was calculated for any energy band in which the pulsar was not detected above the background with a significance of at least 2$\sigma$.\label{spectre_1939}}
\end{center}
\end{figure}

\begin{figure}[ht]
\begin{center}
\includegraphics[scale=0.68]{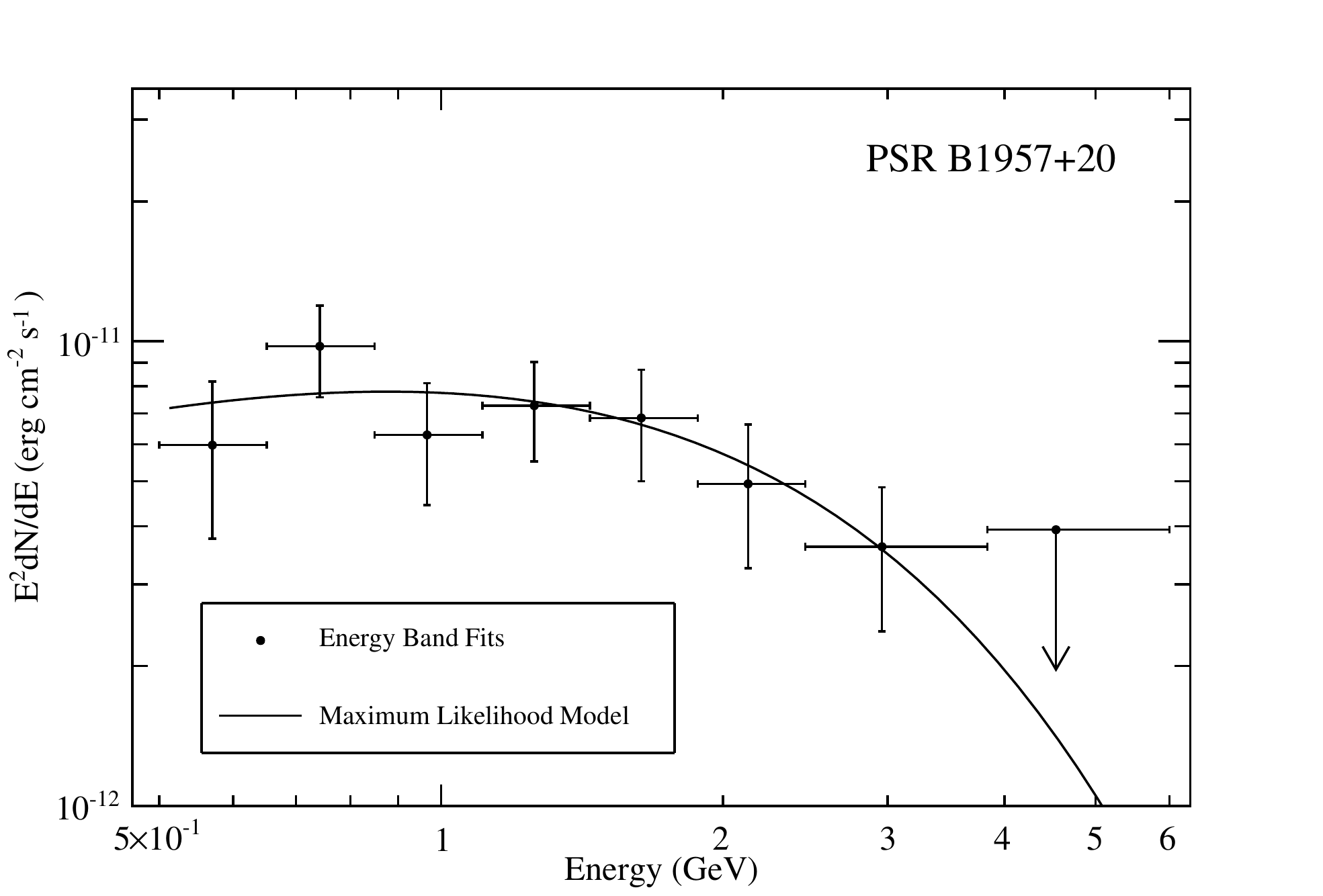}
\caption{Same as Figure \ref{spectre_1939}, for PSR~B1957+20.\label{spectre_1959}}
\end{center}
\end{figure}

The gamma-ray spectra for PSRs B1937+21 and B1957+20 are shown in Figures \ref{spectre_1939} and \ref{spectre_1959}. From Figure \ref{spectre_1959} it can be seen that the full energy range fit for PSR~B1957+20 agrees well with the individual energy band fits. For PSR~B1937+21 the full energy range fit is observed to be systematically below energy band fits between 0.5 and 1 GeV, which may suggest that the actual spectrum is softer than the one fitted using the entire energy range. Maximum likelihood fits over the entire energy range were also performed with the two MSPs modeled with simple power-law spectra ($\beta = 0$) in order to address the significance of measured cutoff energies. Using a likelihood ratio test we find that exponentially cut off power-law models ($\beta = 1$) are preferred at the 2.8 and 3.6$\sigma$ level for PSRs B1937+21 and B1957+20, respectively. Note that a fit of PSR~B1937+21 with a simple power law of the form $N_0 \times \left( E / \mathrm{1 GeV} \right)^{-\Gamma}$ gives $N_0 = (1.75 \pm 0.29) \times 10^{-11}$ cm$^{-2}$ s$^{-1}$ MeV$^{-1}$ and $\Gamma = 3.02 \pm 0.18$. 

To quantify any orbital modulation of the gamma-ray flux of PSR~B1957+20, we constructed a light curve above 0.1 GeV with 10 bins in orbital phase.  We mapped each bin's edges to a series of topocentric times, allowing us to correctly select photons and calculate the exposure to the source.  We then performed a maximum likelihood analysis in which all sources except PSR~B1957+20 were held fixed at their best-fit orbit-averaged values, while the flux of the millisecond pulsar was allowed to vary in each bin. Its spectral shape was held fixed. We found no significant evidence of modulation. To place a limit on the total modulation, we first modeled the orbital modulation with a sinusoid.  By averaging over the zero of phase, we obtained a 95\% confidence upper limit on its peak-to-trough amplitude of 78\%. However, a sinusoid is not an adequate functional form to characterize variability associated with only a limited range of orbital phase, so we also calculated the one-sided 95\% confidence interval for the value of the bins with the maximum and minimum observed flux. To avoid bias from our choice to align the first bin with the zero of orbital phase, we averaged these limits over light curves shifted by 0, $1/30$ and $2/30$. Shifting the data by smaller amounts essentially had no effect on the results. The 95\% limit on the maximum (minimum) flux so averaged is 2.1 (0.1) times the orbit-averaged flux listed in Table \ref{gammaparam}, corresponding to an upper limit on any excursion from the average flux of $\sim 3.4 \times 10^{-8}$ cm$^{-2}$ s$^{-1}$. The latter value provides an upper limit on the emission from shock acceleration, which is consistent with the predictions of models of high-energy emission from colliding winds in massive stars \citep[see e.g.][]{Reimer2006}.

\subsubsection{Gamma-ray pulsations}
\label{gammapuls}

With the full gamma-ray spectral models obtained from the spectral analysis of PSRs B1937+21 and B1957+20, we were able to assign photon probabilities using the \emph{Fermi} ST \emph{gtsrcprob}. We followed the prescriptions described in \citet{Kerr2011weights} and calculated the weighted \emph{H}-test statistics. Selecting events found within 5$^\circ$ from the MSPs and with energies above 0.1 GeV led us to weighted \emph{H}-test parameters of 70.4 for PSR~B1937+21 and 156.1 for PSR~B1957+20, corresponding to pulsation significances of 7.2 and 10.9$\sigma$ respectively, confirming PSRs B1937+21 and B1957+20 as sources of pulsed gamma rays. 

\begin{figure}[ht]
\begin{center}
\includegraphics[scale=0.8]{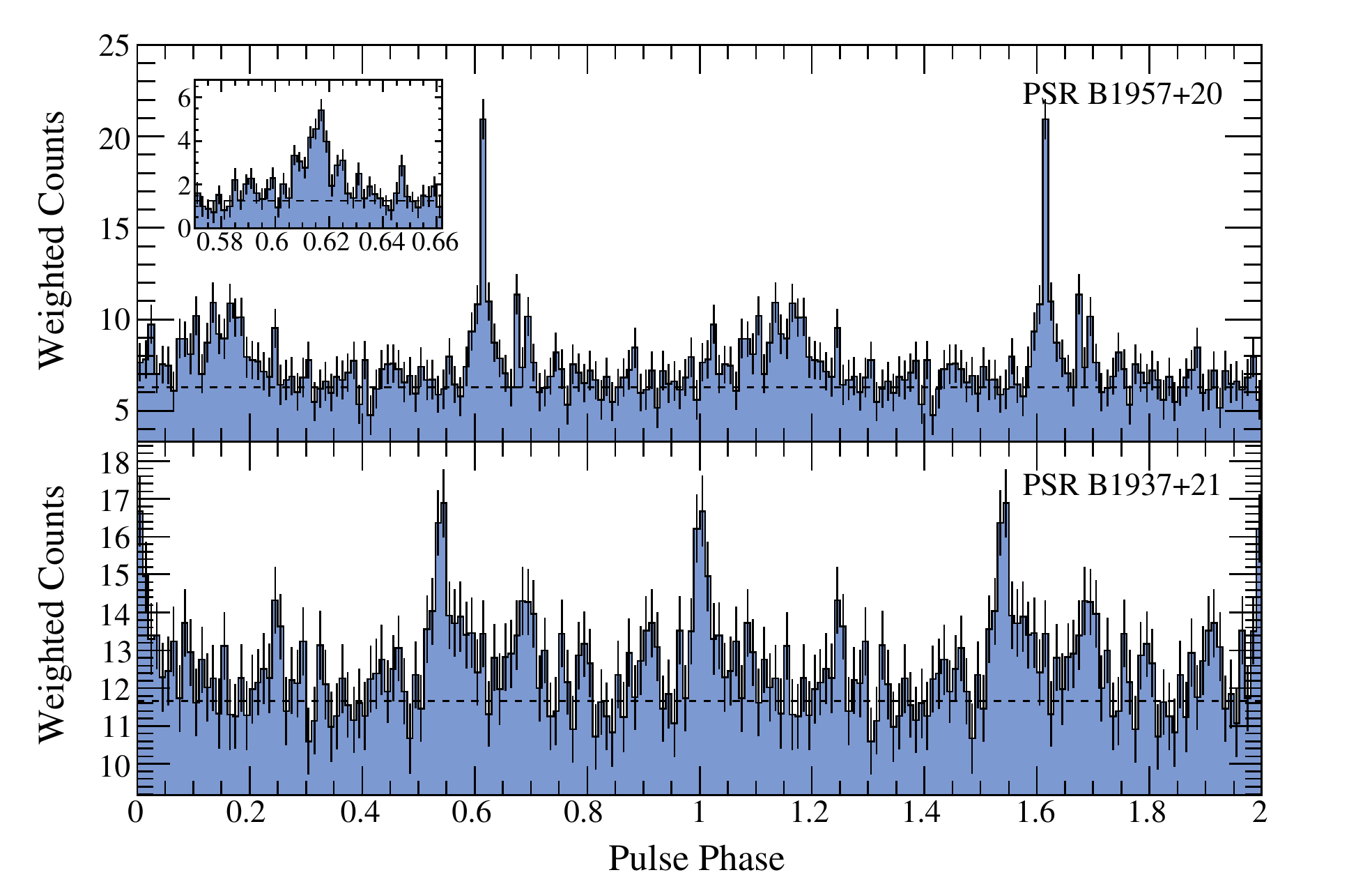}
\caption{Gamma-ray light curves of PSRs B1937+21 (bottom panel) and B1957+20 (top panel) for events recorded by the \emph{Fermi} LAT within 5$^\circ$ from the pulsars, and with energies above 0.1 GeV. These profiles were built by weighting each event by its probability to have been emitted by the pulsars, where the probabilities have been obtained from the spectral analysis of the two MSPs (see \ref{analyse_spectrale} for more details). Two rotations are shown for clarity, and the light curves have 100 bins per rotation. The inset shows the profile of PSR~B1957+20 in the 0.57 -- 0.66 phase range, with 500 bins per rotation or $\sim$ 3 $\mu$s per bin.\label{Wphaso}}
\end{center}
\end{figure}

\begin{figure}[htbp]
\begin{center}
\includegraphics[scale=0.67]{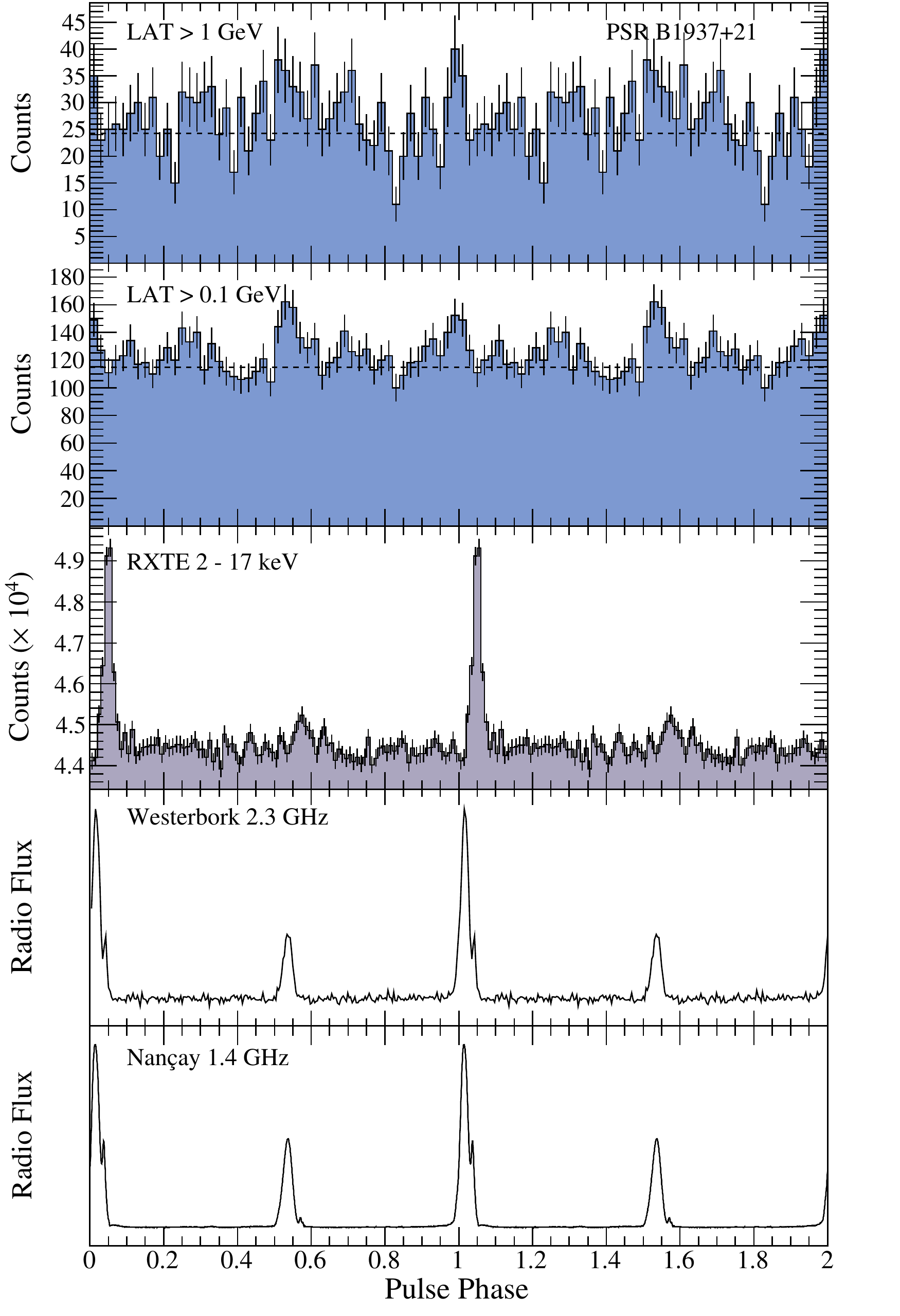}
\caption{Multi-wavelength phase histograms of PSR~B1937+21. The two bottom panels show radio profiles recorded at the Nan\c cay and Westerbork radio telescopes at 1.4 and 2.3 GHz. The middle panel shows an X-ray light curve recorded with \emph{RXTE} between 2 and 17 keV, with 100 bins per rotation. The two top panels show 50-bin gamma-ray light curves recorded with the LAT above 0.1 GeV and 1 GeV. Horizontal dashed lines indicate gamma-ray background levels (see \ref{gammapuls} for details on the determination of these background levels).\label{lc_1939}}
\end{center}
\end{figure}

\begin{figure}[htbp]
\begin{center}
\includegraphics[scale=0.67]{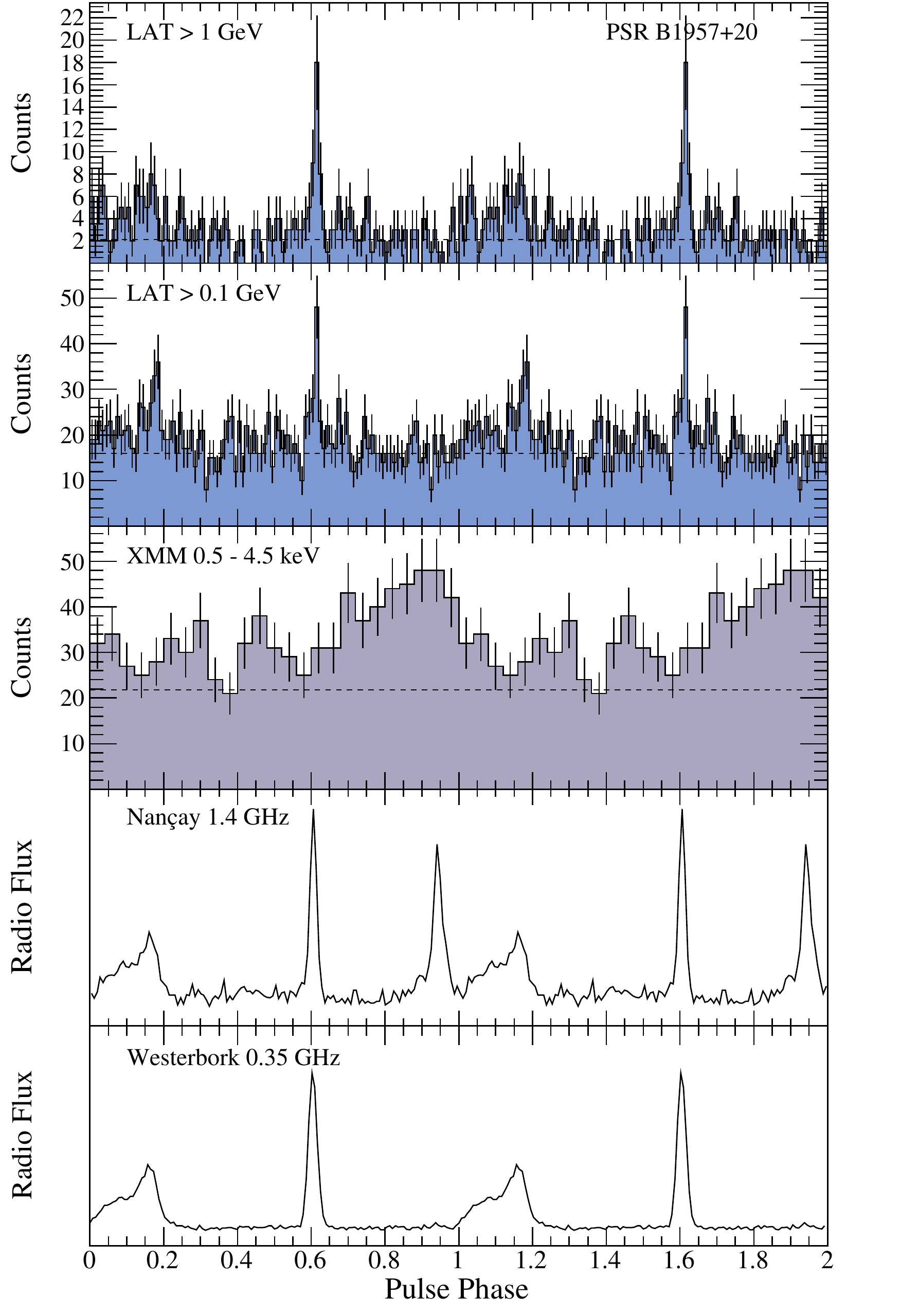}
\caption{Same as Figure \ref{lc_1939}, for PSR~B1957+20. The two bottom panels show radio profiles recorded at the Westerbork and Nan\c cay radio telescopes at 0.35 and 1.4 GHz. The middle panel shows an X-ray profile obtained from an analysis of \emph{XMM-Newton} data between 0.5 and 4.5 keV, with 25 bins per rotation. The background level in this panel was extracted from a similar neighboring region, free from X-ray sources. The two top panels show 100 bin gamma-ray light curves recorded with the LAT above 0.1 GeV and 1 GeV.\label{lc_1959}}
\end{center}
\end{figure}

Figure \ref{Wphaso} shows the weighted light curves for the two MSPs, and as can be seen, PSR~B1937+21 exhibits two main peaks at phases $\sim 0$ and $\sim 0.55$. The error bars were derived by doing a Monte Carlo analysis. In each of 1000 realizations, the photon probabilities calculated with the full spectral model were randomly re-assigned to events in the dataset. For each, a phase histogram (a new weighted light curve) with the same number of bins as in Figure 3 was filled. The uncertainty for a given phase bin was then obtained by calculating the standard deviation of the corresponding phase bin contents in the shuffled light curves. If we denote $s_i = \mathrm{w}_i$ as the probability that a given photon originates from the pulsar, then $b_i = (1 - \mathrm{w}_i)$ gives the probability that the photon is due to background. The total background level $B$ is obtained from the sum, weighted by $b_i$, of the contribution of each photon to the weighted light curve, $s_i$: $B = \sum_i^N s_i \times b_i$. The excess of 12.2 weighted counts above the background level observed at phase $\sim 0.7$ corresponds to a low significance of 1.3$\sigma$. On the other hand, PSR~B1957+20 shows a wide emission component peaking at phase $\sim 0.15$, and a sharp gamma-ray peak at phase $\sim 0.6$. In complement to Figure \ref{Wphaso}, Figures \ref{lc_1939} and \ref{lc_1959} present phase-aligned radio and gamma-ray light curves for PSRs B1937+21 and B1957+20, where the gamma-ray profiles correspond to events found within 0.8$^\circ$ from the pulsars and energies above 0.1 GeV. The absolute phasing in these light curves is such that the maxima of the first Fourier harmonics of the 1.4 GHz Nan\c cay radio profiles transferred back into the time domain define phase 0. Under this convention, the main radio peak and interpulse for PSR~B1937+21 have their maxima at phases $\Phi_{R_1} \sim 0.014$ and $\Phi_{R_2} \sim 0.536$, while for PSR~B1957+20 they fall at phases $\Phi_{R_1} \sim 0.162$ and $\Phi_{R_2} \sim 0.604$. The background levels above 0.1 GeV and 1 GeV shown in Figures \ref{lc_1939} and \ref{lc_1959} were obtained by calculating $B = \sum_i^N (1 - \mathrm{w}_i)$, where $\mathrm{w}_i$ denotes the probability that photon $i$ originates from the pulsar, and $N$ the number of photons in the considered ROI. 

We fitted the gamma-ray emission peaks shown in Figure \ref{Wphaso} using Lorentzian functions. For each peak, the position $\Phi_i$ and the Full Width at Half-Maximum FWHM$_i$ are listed in Table \ref{gammaparam}. Also listed in this Table are radio-to-gamma-ray lags $\delta_i = \Phi_i - \Phi_{R}$, where $\Phi_{R}$ refers to the position of the closest radio peak. Uncertainties quoted in this Table are statistical, and are an order of magnitude larger than absolute phasing uncertainties due to the DM measurement (see Section \ref{ephemerides}). The second gamma-ray peak in PSR~B1957+20's profile is therefore found to be very sharp, with a width of $0.014 \pm 0.007$ at half maximum, or $23 \pm 11$ $\mu$s. We verified the sharpness of this peak by calculating the RMS of weighted phases in the 0.6 to 0.63 range. We obtained a standard deviation of $0.005$ in phase, corresponding to a FWHM of $0.012$, compatible with the FWHM measured in the binned analysis. This peak is the narrowest feature so far observed in a gamma-ray pulsar light curve. 

As can be seen from Figures \ref{lc_1939} and \ref{lc_1959}, PSRs B1937+21 and B1957+20 show nearly aligned radio and gamma-ray emission components, like the Crab pulsar \citep{FermiCrabe} and the millisecond pulsars PSR~J0034$-$0534 \citep{FermiJ0034} and J2214+3000 \citep{Ransom2011}. The radio-to-gamma-ray lags $\delta_i$ for the main and secondary gamma-ray peaks of PSRs B1937+21 and B1957+20 are indeed found to be close to 0, with 1$\sigma$ error bars nearly as large or exceeding the values themselves because of the limited photon sample. In addition, for both pulsars full widths at half-maxima FWHM$_i$ are larger than radio-to-gamma-ray lags $\delta_i$, so that we cannot claim any significant separation. We conclude that for both MSPs the main and secondary emission peaks in gamma rays and radio seem to be produced in similar regions of the pulsars magnetosphere. Interestingly, while all radio peaks of PSR~B1957+20 observed at 0.35 GHz have obvious gamma-ray counterparts, we find no evidence of a gamma-ray peak aligned with the sharp radio peak at phase $\sim 0.95$ observed at 1.4 GHz (see Figure \ref{lc_1959}), suggesting strong spectral dependence of the emission regions and correlation between the gamma-ray and low-frequency radio emission. 

With the firm detection of pulsed gamma-ray emission from PSR~B1957+20, we can now identify the millisecond pulsar as the source powering the \emph{Fermi} LAT First Year Catalog source 1FGL~J1959.6+2047, located 0.5$'$ away \citep{Fermi1FGL}. The energy flux above 0.1 GeV measured for the pulsar in this analysis (see Table \ref{gammaparam}) is in agreement with that of the 1FGL source, of $\sim 2.0 \times 10^{-11}$ erg cm$^{-2}$ s$^{-1}$. In the case of PSR~B1937+21, two gamma-ray sources with no known counterparts are found within a few tens of arc minutes: 1FGL J1938.2+2125c and J1940.1+2209c, with best-fit positions located 21.7' and 35.6' away, respectively. Although the energy flux for PSR B1937+21 above 0.1 GeV is formally consistent with that of the latter 1FGL source of $\sim 3.4 \times 10^{-11}$ erg cm$^{-2}$ s$^{-1}$, it was still detected as a separate gamma-ray source from the pulsar B1937+21, while the former was no longer significant. This, in addition to the relatively large angular separations and the fact that the 1FGL sources are flagged as being possibly spurious or confused with the diffuse emission, indicate that the MSP probably is not formally associated with either of the 1FGL sources. Instead, they may result from the source confusion due to the presence of a weak gamma-ray pulsar in a region of intense background.

\subsection{X-ray analysis}

\subsubsection{PSR B1937+21}

To study the relative alignment of X-ray and gamma-ray emission components, we have re-analyzed the data presented in \citet{Cusumano2003}. Observations of PSR~B1937+21 were made using the Proportional Counter Array \citep[PCA;][]{Jahoda2006} on board the \textit{Rossi X-ray Timing Explorer} (\textit{RXTE}). The PCA is an array of five collimated Xenon/methane multi-anode proportional counter units (PCUs) operating in the 2 -- 60 keV range, with a total effective area of approximately 6500 cm$^2$ and a field of view of $\sim 1^\circ$ FWHM. Data were collected in ``GoodXenon'' mode, which records the arrival time (with 1 $\mu$s resolution) and energy (256 channel resolution) of every un-rejected event. Twenty-one observations taken between 2002 February 21 and 2002 February 28 were downloaded from the HEASARC archive\footnote{http://heasarc.gsfc.nasa.gov/docs/archive.html} and standard selection criteria were applied. Photon arrival times were converted to barycentric dynamical time (TDB) at the solar system barycenter using the JPL DE200 solar system ephemeris with the FITS tool \emph{faxbary}. In order to achieve the best possible absolute timing accuracy available with \emph{RXTE} of $\sim 10$ $\mu$s \citep[e.g.][ including the uncertainty introduced from conversion between TT and TDB]{Rots1998}, the fine clock correction was applied to each event\footnote{ftp://heasarc.gsfc/xte/calib\_data/clock/tdc.dat}. In order to be consistent with the previous analysis of these data described in \citet{Cusumano2003}, we selected photons from all three xenon layers in the energy range 2 -- 17 keV. This analysis produced a total of $4.46\times10^{6}$ photons. Using the parameters and absolute phase information provided in \citet{Cusumano2003}, and correcting an error in the value for the derivative of the DM which was incorrectly given as 2.1 pc cm$^{-3}$ yr$^{-1}$ instead of $2.1\times10^{-3}$ pc cm$^{-3}$ yr$^{-1}$ (M. Kramer, private communication), we folded all of the X-ray photons and created a single pulse profile, shown in Figure \ref{lc_1939}.

Our results are consistent with those of \citet{Cusumano2003}: a fit of the two X-ray emission peaks places the first component at $\phi_{X_1} = 0.0512 \pm 0.0003$ with a FWHM of $0.018 \pm 0.001$ and the second one at $\phi_{X_2} = 0.5752 \pm 0.0004$ with a FWHM of $0.040 \pm 0.012$, in phase units. Error bars are statistical, and do not account for the timing accuracy of \emph{RXTE} of $\sim$ 10 $\mu$s, or 0.006 in phase units. We therefore confirm that the X-ray peaks of PSR~B1937+21 coincide with the giant radio pulse emission and not with the regular radio emission. Additionally, from the positions and widths of the gamma-ray peaks listed in Table \ref{gammaparam}, it seems that X-ray and gamma-ray emissions are misaligned, indicating that they are produced in slightly different regions of the magnetosphere. Note however that the DM variations reported in \citet{Cusumano2003} were not observed in the radio timing analyses made in support of \textit{Fermi} observations (see Section \ref{ephemerides}). In order to exclude any relative phasing issues induced by wrong DM estimates or instrumental issues, we have undertaken the analysis of other archival data supported by multi-frequency radio timing observations spanning over several years. This work will be presented in a future paper.

\subsubsection{PSR B1957+20}

No detection of X-ray pulsations have been reported for PSR~B1957+20 thus far \citep[see][for a recent analysis]{Huang2007}. However, it must be noted that folding the gamma-ray data for PSR~B1957+20 with the pulsar ephemeris used in the latter paper does not reveal gamma-ray pulsations. While this shows that the ATNF Catalogue timing solution used in \citet{Huang2007} is invalid for the \emph{Fermi} data taken after August 2008 it does not prove that the ephemeris was invalid for the \emph{XMM-Newton} data taken in 2004. This motivated a re-analysis of the same X-ray data, using the ephemeris presented in Section \ref{ephemerides}.

PSR~B1957+20 was observed with {\em XMM-Newton} on 2004 October 31 and 2004 November 1. The observations with the three European Photon Imaging Cameras (EPIC) MOS1, MOS2 (imaging mode) and pn (timing mode) spanned approximately 30 ks. These instruments were operated in full-frame mode with a thin filter. We used Version 10.0 of the \emph{XMM-Newton} Science Analysis Software\footnote{http://xmm.esac.esa.int/sas/} to reduce and analyze the X-ray data. The EPIC Observation Data Files (ODFs) were reduced using the \emph{emproc/epproc} scripts (for MOS and pn respectively) along with the most recent Calibration Files (CCFs) at the time of the reduction. We applied standard filtering procedures \citep{Webb2004}, which includes the removal of strong background periods caused by soft photon flares, and considered the data between 0.3 and 10 keV for the spectral analysis. We restricted the timing analysis to the 0.5 -- 4.5 keV band as this was found to have the best signal-to-noise. We removed the events contained in energy ranges affected by internal background caused by the X-ray fluorescence of satellite material exposed to cosmic rays\footnote{http://xmm2.esac.esa.int/docs/documents/CAL-TN-0018.pdf}.

We extracted MOS spectra from circular regions of radii 45$''$ centered on the radio position of the pulsar, which allowed us to collect about 90\% of all detected source counts and optimize the signal to noise ratio. In the pn timing mode, the central CCD is read out continuously at high speed, and the collected events are condensed into a one dimensional pixel array. The source event file is therefore extracted  by selecting a rectangular region around the expected position of the pulsar. We extracted MOS and pn background photons from neighboring regions free of X-ray sources, and generated instrumental response files with the \emph{RMFGEN/ARFGEN} tasks.

We fitted the combined MOS1/2 and pn spectra with \emph{Xspec} Version 12.6.0\footnote{http://heasarc.gsfc.nasa.gov/xanadu/xspec/}. Using a simple power-law model yielded spectral results that are consistent with those of \citet{Huang2007}, with a photon index $\Gamma = 2.37^{+0.57}_{-0.29}$ ($\chi^2 = 0.70$ for 27 d.o.f., errors are 90\%). The fitted column absorption along the line-of-sight $N_{H}$ is $15.7^{+9.6}_{-9.0} \times 10^{20}$ cm$^{-2}$, and is consistent with the total Galactic HI column density in the direction of the pulsar given by the HEASARC Tool $N_{H}$\footnote{http://heasarc.gsfc.nasa.gov/cgi-bin/Tools/w3nh/w3nh.pl} ($\sim 30 \times 10^{20}$ cm$^{-2}$). The unabsorbed X-ray flux in the band 0.2 -- 10 keV is F$_{X} = \left( 9.7 \pm 0.8 \right) \times 10^{-14}$ erg cm$^{-2}$ s$^{-1}$. Using the NE2001 distance of 2.5 kpc, the X-ray luminosity is found to be $L_{X} = \left( 7.2 \pm 0.6 \right) \times 10^{31}$ erg s$^{-1}$. Assuming the emission is produced by the pulsar, we estimate that the efficiency of the conversion of spin-down energy loss $\dot E$ into X-ray emission is $\eta_X = L_X / \dot E \sim 9.7  \times 10^{-4}$. These values are higher than the ones presented by \citet{Huang2007}, who considered a distance to the pulsar of 1.5 kpc. However, with the lower limit on the distance of $d \gtrsim 2$ kpc placed by \citet{vanKerkwijk2011}, the X-ray luminosity and efficiency values measured in this analysis are favored. 

The relative timing accuracy of \emph{XMM-Newton} has been proven to be better than $10^{-8}$, while the absolute timing accuracy is $\sim 53$ $\mu$s \citep{MartinCarrillo2011}, allowing to search for pulsations in the X-ray data using the pn source event file extracted as described above. The event times were converted to TDB using the task \emph{barycen} Version 1.18 and the JPL DE405 solar system ephemeris, and phase-folded using \textsc{Tempo2} and the ephemeris described in Section \ref{ephemerides}. This timing solution however does not formally cover the X-ray data considered here, taken four years before the beginning of the radio timing dataset. Millisecond pulsars are generally stable rotators, therefore it is reasonable to extrapolate the ephemeris to our observations in a similar way to \citet{Bogdanov2010}. \citet{Arzoumanian1994} measured a DM for PSR~B1957+20 of 29.1168 $\pm$ 0.0007 pc cm$^{-3}$ at 48196 MJD. We can therefore safely assume that the DM value at the time of the \emph{XMM-Newton} observations was comprised between the latter value of \citet{Arzoumanian1994} and our DM measurement of 29.12644 $\pm$ 0.00029 pc cm$^{-3}$ at 54550 MJD. The difference in DM between 48196 and 54550 MJD of $\sim 9.6 \times 10^{-3}$ pc cm$^{-3}$ introduces an uncertainty on the relative phasing of \emph{XMM-Newton} data and 1.4 GHz radio data of $\sim$ 20 $\mu$s. This relatively small value indicates that the uncertainty on the DM at the time of the X-ray observations should not affect our timing analysis of the \emph{XMM-Newton} data significantly. Nevertheless, our analysis could be affected by the instability of the pulsar's rotational frequency or orbital movement. 

The X-ray phase histogram is shown in Figure \ref{lc_1959}. The pulse profile gathers 854 events taken between 0.5 and 4.5 keV, of which $\sim$ 33\% are estimated to come from the pulsar. The $H$-test parameter for this set of events is 24, which corresponds to a pulsation significance of $\sim 4 \sigma$. Assuming the ephemeris was accurate at the time of the \emph{XMM} observations, it hence seems that PSR~B1957+20 emits pulsed X-rays. The background level shown in Figure \ref{lc_1959} was calculated by extracting a similar neighboring region free from X-ray sources, as mentioned above. We fitted the X-ray peak with a Gaussian and found that it occurs at $\phi_X$ = 0.86 $\pm$ 0.03, with a FWHM of 0.32 $\pm$ 0.03. It is therefore found to be offset from the aligned radio and gamma-ray peaks by at least 150 $\mu$s. From the absolute timing uncertainty of $\sim 73$ $\mu$s $\sim 0.05$ in phase induced by \emph{XMM-Newton}'s intrinsic absolute timing capability and the uncertainty on the DM, it seems that the X-ray peak is separated from the phase-aligned radio and the gamma-ray emission peaks. We cannot exclude however drifting induced by non-optimal rotational and binary parameters, causing the pulse phases to be offset from the actual values. A longer X-ray observation of the MSP, analyzed with a contemporaneous timing solution would be preferable for confirming the phase and shape of the peaks.

\section{Discussion}

\subsection{Light curve modeling}
\label{modeling}

The first eight MSPs discovered using the \textit{Fermi} LAT \citep{Fermi8MSPs} exhibited non-zero lags between their respective radio and gamma-ray light curves, and have been modeled using either standard two-pole caustic (TPC) and outer gap (OG) geometries, or pair-starved polar cap (PSPC) models \citep{Venter2009}, or alternatively, an annular gap model \citep{Du2010}. Light curve modeling using a force-free magnetospheric geometry may present a further possibility \citep{Bai2010_2,Contopoulos2010}.  The common feature of all these models is that the gamma-ray emission originates in the outer magnetosphere and not near the polar caps. The discovery of radio and gamma-ray light curve peaks in close alignment for J0034$-$0534 \citep{FermiJ0034} yielded the first example of an MSP with such phase-aligned peaks, previously only seen in the Crab pulsar. The newly-discovered PSR J2214+3000 \citep{Ransom2011} also shows aligned radio and gamma-ray emission peaks. As can be seen in Figures \ref{lc_1939} and \ref{lc_1959} showing multi-wavelength light curves for PSRs B1937+21 and B1957+20, we now have a class of MSPs with phase-aligned low and high-energy pulsations, in addition to PSRs J0034$-$0534 and J2214+3000. 

Light curve modeling efforts for PSRs B1937+21 and B1957+20 are guided by the following key observed features: (i) gamma-ray and radio peak phase alignment within the statistical uncertainties; (ii) gamma-ray profiles that have multiple sharp peaks and possible increased complexity if additional low-level features become more significant with accumulated photon statistics. 

\begin{figure}
\begin{center}
\includegraphics[scale=0.67]{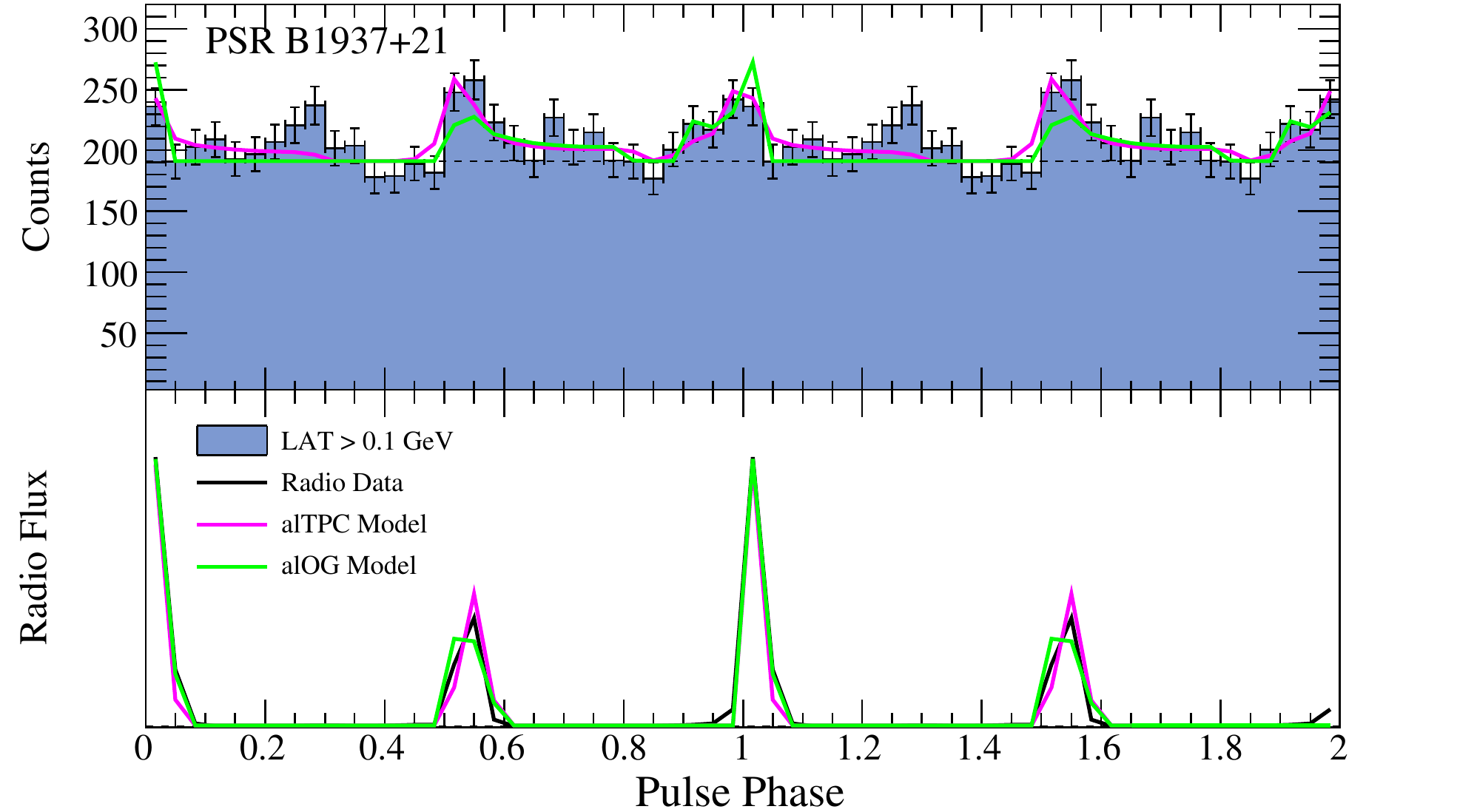}
\caption{Top: gamma-ray data and modeled light curves for PSR~B1937+21 with 30 bins per rotation. Bottom: Nan\c cay 1.4 GHz radio profile and modeled light curves. See Table \ref{fitting_results} for the best-fit parameters. \label{model_1939}}
\end{center}
\end{figure}

\begin{figure}
\begin{center}
\includegraphics[scale=0.67]{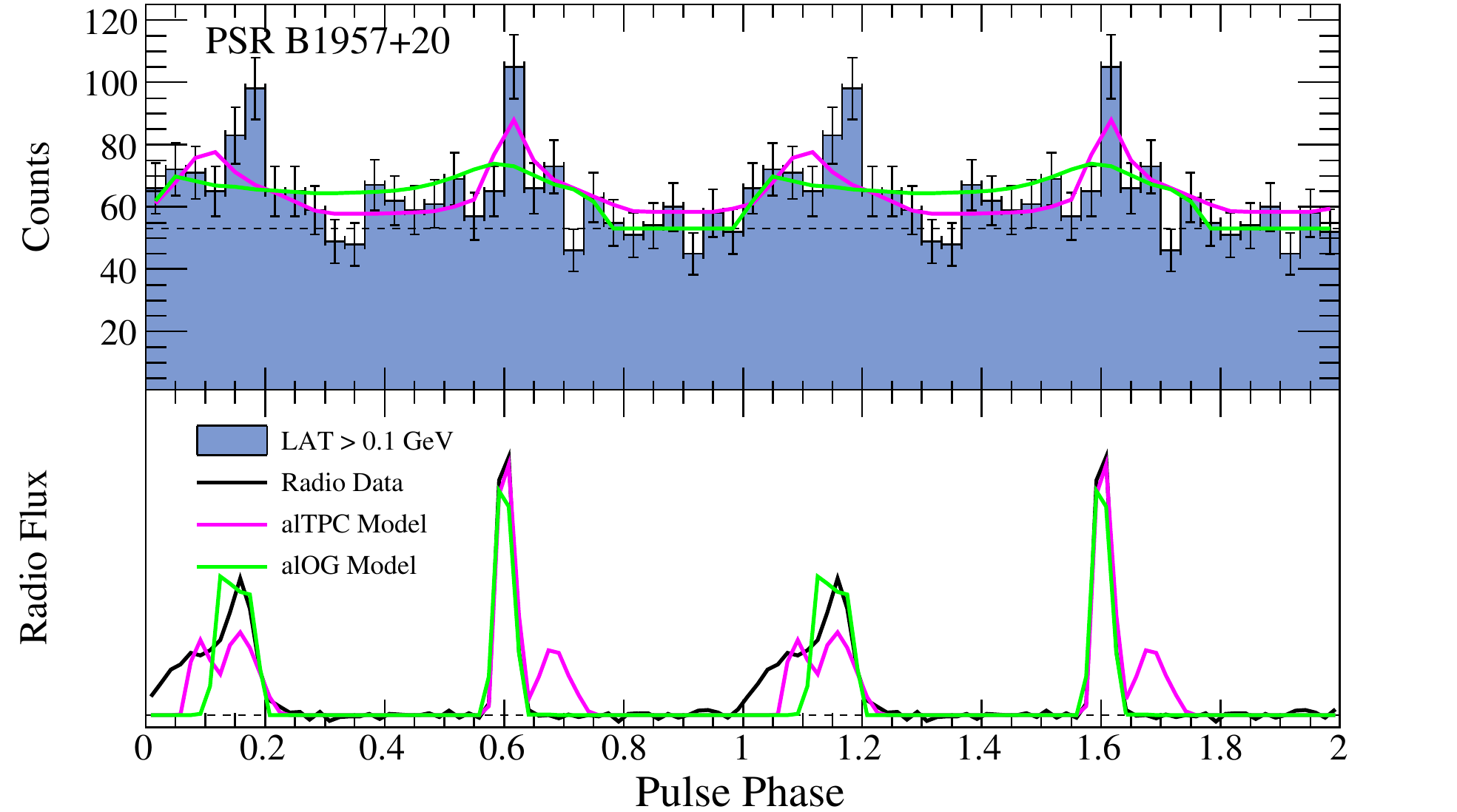}
\caption{Same as Figure \ref{model_1939}, for PSR~B1957+20. The bottom panel shows the Westerbork 0.35 GHz radio profile. \label{model_1959}}
\end{center}
\end{figure}

As described in \citet{FermiJ0034}, the alignment of gamma-ray and radio peaks suggests co-location of gamma-ray and radio emission regions in the pulsar magnetosphere. We obtain reasonable fits for both gamma-ray and radio light curves in the context of ``altitude-limited'' TPC (alTPC) and OG (alOG) models (Figures \ref{model_1939} and \ref{model_1959}).  These models assume that the radio emission is emitted within a range of altitudes relative to the light cylinder (at radius $R_{LC} = c P / 2 \pi$), along the last open field lines, with the same geometry as the gamma-ray emission except that the emission is limited to a smaller range of altitudes. Presently, high-altitude emission seems to be preferred to low-altitude emission, although the latter cannot be ruled out \citep{Venter2011}. Thus, both the gamma-ray and radio peaks are caustics, resulting from phase-bunching of photons due to relativistic effects associated with high corotation velocities in the outer magnetosphere \citep{Dyks2004,Romani1995}. For most viewing geometries, caustic radio emission leads to large amounts of depolarization due to mixing of emission from different altitudes; however, orthogonal configurations with viewing angles near 90$^{\circ}$ result in less depolarization and thus a careful study of the expected polarization properties in these models is warranted.

In order to statistically pick the best-fit parameters for the alTPC and alOG models for the two MSPs considered here we have developed a Markov chain Monte Carlo (MCMC) maximum likelihood procedure \citep{Johnson2011b}. An MCMC involves taking random steps in parameter space and accepting a step based on the likelihood ratio with respect to the previous step \citep{Hastings1970}. The gamma-ray light curves are fitted using Poisson likelihood while the radio light curves are fitted using a $\chi^{2}$ statistic and the two values are combined.  For a given parameter state the likelihood value is calculated by independently optimizing the radio and gamma-ray model normalizations using the \emph{scipy} python module\footnote{See http://docs.scipy.org/doc/ for documentation.} and the \emph{scipy.optimize.fmin\_l\_fbgs\_b} multi-variate bound optimizer \citep{Zhu1997}. The likelihood surfaces can be very multi-modal which can lead to poor mixing of the chain and slow convergence.  Therefore, we have implemented small-world chain steps \citep{Guan2006} and simulated annealing \citep{Marinari1992} to speed up the convergence and ensure that the MCMC fully explores the parameter space and does not get stuck in a local maximum.  We verify that our chains have converged using the criteria proposed by \citet{Gelman1992}.

In order to balance the gamma-ray and radio contributions to the likelihood, we have chosen to use an uncertainty for the radio intensity which is equal to the average, relative gamma-ray uncertainty in the on-peak region times the maximum radio value. The best-fit results can be strongly affected by the uncertainty which is chosen for the radio profile; in particular, a smaller uncertainty will decrease the overall likelihood and can, in some cases, lead to a different best-fit geometry which favors the radio light curve more strongly.  When varying the radio uncertainty by a factor of 2, the best-fit $\alpha$ and $\zeta$ values of PSR J1939+2134 were found to vary by $\leq7^{\circ}$.  The best-fit geometry of PSR J1959+2048 was found to be more sensitive to changes in the radio uncertainty with either the best-fit $\alpha$ or $\zeta$ value changing by $\sim35^{\circ}$ while the other parameter changed by $\lesssim15^{\circ}$.

Our simulations have a resolution of 1$^\circ$ in both $\alpha$ and $\zeta$, 0.05 in gap width ($w$, normalized to the polar cap angle $\theta_{pc} \sim\sqrt{R_{NS}/R_{LC}}$), and $0.1 R_{LC}$ for the minimum and maximum emission altitudes.  The MCMC explores viewing geometries for $\alpha$ from 1$^{\circ}$ to 90$^{\circ}$ and for $\zeta$ from 0$^{\circ}$ to 89$^{\circ}$, as shown in Figure \ref{RVM_chi2}.  We expect that going beyond 90$^\circ$ should produce the same profiles, simply shifted in phase by 180$^\circ$. Additionally, for the gamma-ray models we restrict the maximum emission altitudes to be $\geq 0.7 R_{LC}$, as high-altitude emission near the light cylinder is important for producing the correct pulse shapes. In addition to predicting the best-fit model parameters the simulations also provide numerical estimates of the beaming correction factor ($f_{\Omega}$) as described in \citet{Watters2009} and \citet{Venter2009}. For a given $\alpha$ and set of emission parameters (i.e., gap width and emission altitudes), the $f_{\Omega}$ factor is calculated by collecting the emission at a specified $\zeta$ and comparing it to the total emission for all viewing angles.

The radio and gamma-ray profiles of PSRs B1937+21 and B1957+20 were reproduced using the alTPC and alOG geometries with the best-fit parameters listed in Table \ref{fitting_results}, the uncertainties on the model parameters were derived from a likelihood profile scan. In this Table, $R_{NS}$ denotes the neutron star radius, and $R_{NCS}$ is the field-line-dependent altitude of the null-charge surface, defined by $\mathbf{\Omega} \cdot \mathbf{B} = 0$. Best-fit models with infinitely thin gap widths ($w = 0$) do not represent the truth as a zero-width gap is unphysical. However, they indicate that the best gap width is somewhere between 0 and 0.05 and the best-fit value of 0 is chosen only as a result of the resolution of our simulations. For PSR~B1937+21, the fits yield large $\alpha$ and $\zeta$ values, which reinforces the idea that this MSP may be a nearly orthogonal rotator, as is suggested by the fact that the observed radio interpulse lags the main radio peak by approximately half a rotation. Recent observations using the Parkes telescope indicate that PSR~B1937+21 may have a large duty cycle of $\sim$ 80\% \citep{Yan2011}. The two main peaks separated by 172$^\circ$ remain the main features of the profile, while off-peak features at $\sim$ 0.5\% of the peak intensity are seen. While the current radio models reproduce the two main peaks very well, it may be worthwhile to model the extended low-level off-peak emission as well in future, as these features may result from emission regions below the null charge surface. If this is true, emission from both poles is implied, favoring a TPC geometry.

The alTPC and alOG models also yield similar values of the $\alpha$ and $\zeta$ angles for PSR~B1957+20, preferring moderate values of $\alpha$ and $\zeta$ near 90$^{\circ}$. The spin-up of millisecond pulsars by mass transfer from a companion star naturally leads to aligned orbital and rotation axes \citep{ThorsettStinebring1990}. Since the pulsar shows eclipses \citep{Fruchter1988}, the line-of-sight is nearly parallel to the plane of the orbit, which indicates that a predicted observer angle $\zeta$ close to 90$^\circ$ is preferred.  However, neither model is able to successfully reproduce the narrow gamma-ray peaks well and, as can be seen in Table~\ref{fitting_results}, $\alpha$ is not well constrained, particularly for the alOG model.  With more statistics the gamma-ray peaks will be more important to the likelihood and the fits may improve.

\subsection{Constraints on orientation angles from radio data}

In contrast with most other MSPs, the radio polarization profile of PSR~B1937+21 is simple, with a distinct flat polarization angle (PA) swing \citep[see e.g.][]{Stairs1999}. This suggests that the orientation angles of the pulsar can be inferred by fitting the polarization data with the rotating vector model \citep[RVM,][]{Radhakrishnan1969}, accounting for the aberration effect bending the emission beam in the co-rotational frame of the pulsar forward with respect to the rotating frame, and resulting in a delay of the PA swing respective to the intensity profile by $4 r/R_{LC}$ radians, where $r$ is the emission radius \citep{Blaskiewicz1991}. We therefore attempted to model the radio emission geometry, using the 0.6 and 1.4 GHz polarization data from \citet{Stairs1999}, obtained from the EPN database\footnote{http://www.mpifr-bonn.mpg.de/div/pulsar/data/browser.html}. The degree of linear polarization of the main peak is fairly high ($> 50 \%$) while that of the secondary peak is $10 \%$. A PA jump due to orthogonal polarization modes (OPM) is identified in the main peak at 1.4 GHz, and in the secondary peak at 0.6 GHz. The continuity of the OPM jumps and the fact that they are not exactly $90^{\circ}$ apart may imply that the PA swings are mildly affected by scattering in the interstellar medium \citep{Karastergiou2009}. 

\begin{figure}
\begin{center}
\includegraphics[scale=0.6]{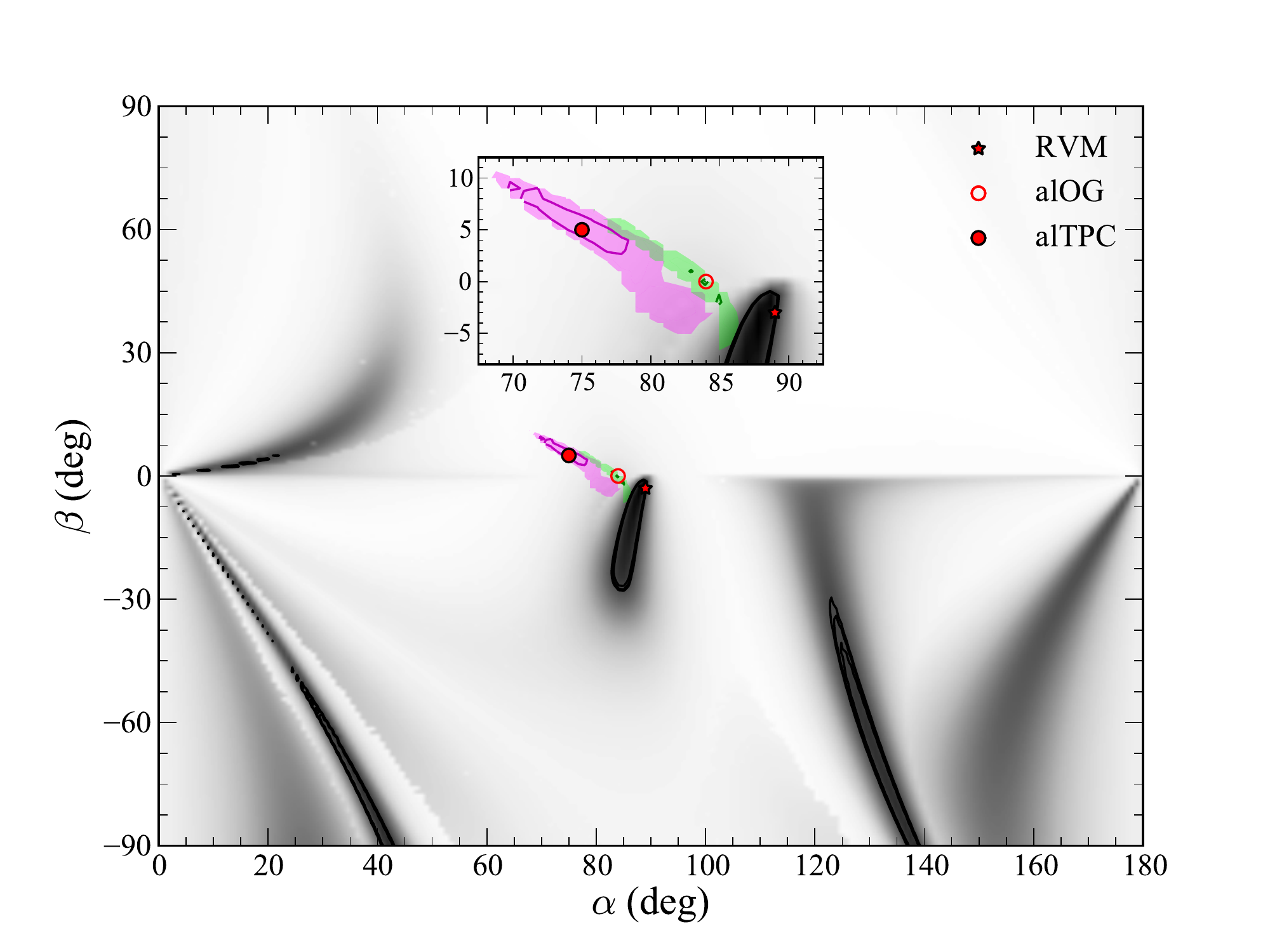}
\caption{Plot of the impact angle $\beta = \zeta - \alpha$, vs. the magnetic inclination angle $\alpha$ for PSR~B1937+21. The $\chi^2$ map of the rotating vector model (RVM) fit of radio polarization data for PSR~B1937+21 is shown in gray scale, with contour levels at 1, 2 and 3$\sigma$ in black. Magenta and green shaded regions indicate the 3$\sigma$ contour levels of the simultaneous radio and gamma-ray light curve modeling under the alTPC and alOG geometries, while the solid lines show the 1$\sigma$ contour levels. The symbols indicate the best pulsar orientation angles obtained from the RVM, alTPC and alOG fits.\label{RVM_chi2}}
\end{center}
\end{figure}

The PA profiles at 0.6 and 1.4 GHz corrected for the OPM jumps are very similar. In addition, RVM fits of the two PA profiles give consistent results. We therefore merged the two profiles into a single PA swing. Figure \ref{RVM_chi2} shows the results of the RVM fit to the merged PA swing. The best fit of the data is obtained for $\alpha$ and $\beta$ angles of 89$^\circ$ and $-3^\circ$ respectively, where $\beta = \zeta - \alpha$ is the angle between the magnetic axis and the line of sight. Also shown in the plot are the pulsar orientation angles obtained from the modeling of the radio and gamma-ray profiles under the alTPC and alOG geometries. As can be seen from Figure \ref{RVM_chi2}, the best-fit configuration is found in the vicinity of the angles obtained from the modeling. Both the radio and gamma-ray light curve modeling and the analysis of the polarization data therefore support values of the $\alpha$ and $\zeta$ angles close to 90$^\circ$, corresponding to an orthogonal configuration. 

In addition to inferring pulsar geometric angles, the RVM fit also allowed us to estimate the altitude of the radio emission, and compare it to the modeled values listed in Table \ref{RVM_chi2}. We find that the magnetic axis is located at $\sim150^{\circ}$ after the secondary peak ($\sim 0.42$ in phase), corresponding to a radio emission altitude of 0.65 $R_{LC}$. This value is consistent with the emission altitude found from the modeling under both alOG and alTPC geometries, confirming that the radio emission seems to be produced at high altitudes above the neutron star surface. We note that the aberration treatment presented by \citet{Blaskiewicz1991} and later by \citet{Hibschman2001} and \citet{Dyks2008} is only a first-order approximation that may need correcting, in particular at high altitudes. However, the consistency of the results obtained from the RVM fitting and the light curve modeling gives us some confidence in suggesting that both radio and gamma-ray emission originate from the outer magnetosphere, which for PSR~B1937+21 is compact anyway. 

On the other hand, radio polarization data for PSR~B1957+20 could not be used to constrain the pulsar orientation angles. This pulsar displays less than 2\% linear polarization though some evidence exists for sign-changing, circular polarization through both peaks \citep{ThorsettStinebring1990}.  Observations well away from the eclipsing phase have confirmed that the lack of polarization is not due to interactions with the companion. This lack of linear polarization is consistent with radio emission of a caustic nature. Note that similar behavior is observed for PSR J0034$-$0534 \citep{Stairs1999} which has also been fit with the alTPC and alOG models \citep{FermiJ0034,Venter2011}. Future modeling efforts should be able to reproduce the weak polarization of caustic peaks. 

\subsection{Gamma-ray luminosities}
\label{luminosities}

Knowing the pulsar distance $d$ and the gamma-ray energy flux $G$ measured using the LAT above 0.1 GeV, one can derive the gamma-ray luminosities $L_\gamma = 4 \pi f_\Omega G d^2$ and efficiencies of conversion of spin-down energy into gamma-ray emission $\eta = L_\gamma / \dot E$ of PSRs B1937+21 and B1957+20. In these expressions, $f_\Omega$ is the correction factor depending on the viewing geometry discussed in Section \ref{modeling}. Table \ref{gammaparam} lists values of $L_\gamma$ and $\eta$ under the assumption that $f_\Omega = 1$. The geometrical correction factors $f_\Omega$ obtained from the modeling of radio and gamma-ray light curves with alTPC and alOG geometries are listed in Table \ref{fitting_results}, as well as the corresponding corrected gamma-ray luminosity and efficiency values. 

With gamma-ray luminosity estimates on the order of $\sim 2.5 \times 10^{35}$ erg s$^{-1}$, PSR~B1937+21 does not stand out from gamma-ray pulsars with comparable $\dot E$ values \citep[see e.g.][]{FermiPSRCatalog}. Nevertheless, the gamma-ray luminosity above 0.1 GeV inferred from using the NE2001 distance of $3.6 \pm 1.4$ kpc is $L_\gamma / f_\Omega = (5.63 \pm 3.95 \pm 3.53) \times 10^{34}$ erg s$^{-1}$ (where the first uncertainty quoted is statistical and the second is systematic). This value is consistent with those of pulsars with similar $\dot E$ values within uncertainties, and therefore neither parallax distance estimates nor distances based on Galactic electron density models are favored with the current gamma-ray analysis and knowledge of the relationship between $L_\gamma$ and $\dot E$.

Similarly, the gamma-ray luminosities measured for PSR~B1957+20 are consistent with those of pulsars with comparable $\dot E$ values to within uncertainties, despite the fact that the $f_\Omega$ correction factors obtained from the modeling are very different between the two models. We therefore cannot constrain the distance of 2.5 kpc inferred from the NE2001 model of Galactic electron density. Nevertheless the low gamma-ray efficiency inferred from the alOG model is very similar to those of the bulk of MSPs \citep{Fermi8MSPs}. This fact, in addition to the inferred $\zeta$ being close to 90$^\circ$ as suggested from the occurrence of radio eclipses and the slightly better likelihood parameter suggests that the altitude-limited OG model is preferred over the TPC model for PSR~B1957+20.

\section{Summary and conclusions}

We have reported the discovery of phase-aligned radio and gamma-ray light curve peaks for the millisecond pulsars PSRs B1937+21 (the ``first'' MSP) and B1957+20 (the first black widow system). This adds two new members to the class of MSPs exhibiting the phenomenon of such phase-aligned peaks. 

The fact that we find reasonable (but possibly non-unique) radio and gamma-ray light curve fits implies that the geometric caustic models still provide an adequate description for this new class of MSPs. As noted before \citep{Venter2009}, the sharp peaks, coupled with the caustic fits, imply copious pair production in the MSP magnetospheres, in contrast to earlier expectations. Cascades of electron-positron secondaries are needed to set up and sustain the TPC / OG geometries which reproduce the salient features of the light curves. PSR~B1937+21 is very near the death line for screening by pairs, but PSR~B1957+20 lies well below the line and is not expected to produce enough pairs for screening in conventional polar cap models assuming dipolar magnetic fields \citep{Harding2005}. In addition, these geometric models provide a framework to constrain the emission altitudes of the gamma-ray and radio photons, as well as the beaming and inclination-observer geometry of the MSPs. Lastly, observations of the radio emission at different frequencies, as well as energy-dependent light curve modeling, may provide the opportunity to learn more about the radius-to-frequency mapping of the radio, its connection with the gamma-ray radiation, and ultimately the mechanism for the generation of the radio emission.

Our analysis of the \emph{RXTE} X-ray data for PSR~B1937+21 yielded results that are consistent with those of \citet{Cusumano2003}. The X-ray light curve consists of two peaks lagging the regular radio emission by a small amount, but in close alignment with the giant radio pulse emission. X-ray pulses also are not formally aligned with the gamma-ray emission seen with the LAT. Detailed modeling of the X-ray emission geometry for this MSP would help understand the misalignment with the regular radio emission and the gamma-ray emission. Nevertheless, the sharpness of the X-ray peaks and their proximity to the outer magnetospheric radio and gamma-ray emissions suggest that the X-ray emission from PSR~B1937+21 also takes place at high altitude in its magnetosphere, which is supported by the non-thermal nature of the emission \citep{Cusumano2003,Nicastro2004}. 

On the other hand, we have found evidence for X-ray pulsed emission from PSR~B1957+20 using \emph{XMM-Newton} data, for the first time. The current timing analysis suffers from potential phase drifting due to the fact that our radio timing solution does not cover the X-ray data epoch. Nevertheless, the relatively large spin-down luminosity $\dot E$ of this pulsar of $7.5 \times 10^{34}$ erg s$^{-1}$, comparable to that of other X-ray emitting MSPs, makes it a good candidate for pulsed X-ray emission. Besides, this spin-down luminosity is characteristic of non-thermally emitting MSPs, which is supported by the spectral analysis of the \emph{XMM-Newton} data. If PSR~B1957+20 does emit pulsed X-rays with a non-thermal spectrum, then the X-ray emission from this MSP is expected to align with the radio emission, as is the case for PSRs B1937+21 and J0218+4232 \citep{Kuiper2002}. Is is therefore important to observe this pulsar in X-rays again, with a radio timing solution covering the X-ray data and ensuring the validity of the absolute phasing. 

\acknowledgments

The \textit{Fermi} LAT Collaboration acknowledges generous ongoing support from a number of agencies and institutes that have supported both the development and the operation of the LAT as well as scientific data analysis. These include the National Aeronautics and Space Administration and the Department of Energy in the United States, the Commissariat \`a l'Energie Atomique and the Centre National de la Recherche Scientifique / Institut National de Physique Nucl\'eaire et de Physique des Particules in France, the Agenzia Spaziale Italiana and the Istituto Nazionale di Fisica Nucleare in Italy, the Ministry of Education, Culture, Sports, Science and Technology (MEXT), High Energy Accelerator Research Organization (KEK) and Japan Aerospace Exploration Agency (JAXA) in Japan, and the K.~A.~Wallenberg Foundation, the Swedish Research Council and the Swedish National Space Board in Sweden.

Additional support for science analysis during the operations phase is gratefully acknowledged from the Istituto Nazionale di Astrofisica in Italy and the Centre National d'\'Etudes Spatiales in France.

The Nan\c cay Radio Observatory is operated by the Paris Observatory, associated with the French Centre National de la Recherche Scientifique (CNRS). The Westerbork Synthesis Radio Telescope is operated by Netherlands Foundation for Radio Astronomy, ASTRON.

\bibliographystyle{apj}

\bibliography{biblio}

\clearpage

\begin{table}
\begin{center}
\begin{small}
\caption{Properties of PSRs B1937+21 and B1957+20. Values in parentheses indicate the 1$\sigma$ uncertainties on the last digit quoted. For~PSR B1937+21, $P$ and $\dot P$ and $\mu_T$  values are taken from \citet{Cognard1995}, while values for PSR~B1957+20 are taken from \citet{Arzoumanian1994}. These value are given at epochs MJD 47899.5 and 48196 in TDB units, respectively. Distances $d$ are derived from timing parallax measurements from \citet{Verbiest2009} for PSR~B1937+21, and from the NE2001 model of Galactic electron density \citep{NE2001} for PSR~B1957+20. The total apparent proper motion for PSR~B1937+21 is small, making the Shklovskii contribution to the period derivative value \citep{Shklovskii1970} almost negligible. The last three parameters were calculated using the intrinsic spin-down rate $\dot P_{corr}$, corrected for the Shklovskii effect. \label{proprietes}}
\begin{tabular}{lcc}
\tableline\tableline
Parameter & PSR~B1937+21 & PSR~B1957+20 \\
\tableline
Galactic longitude, $l$ (deg) & 57.51 & 59.20 \\
Galactic latitude, $b$ (deg) & $-0.29$ & $-4.70$ \\
Pulsar period, $P$ (ms) & 1.557806472448817(3) & 1.60740168480632(3) \\
Apparent period derivative, $\dot P$ ($10^{-21}$) & 105.1212(2) & 16.8515(9) \\
Transverse proper motion $\mu_{T}$ (mas yr$^{-1}$) & 0.80(2) & 30.4(6) \\
Distance $d$ (kpc) & 7.7 $\pm$ 3.8 & 2.5 $\pm$ 1.0 \\
Corrected period derivative, $\dot P_{corr}$ ($10^{-21}$) & 105.10 $\pm$ 0.01 &  7.85 $\pm$ 3.61 \\
Spin-down luminosity, $\dot E$ ($10^{34}$ erg/s) & 109.76 $\pm$ 0.01 & 7.48 $\pm$ 3.43 \\
Surface magnetic field, $B_{surf}$ ($10^{8}$ G) & 4.0946 $\pm$ 0.0003 & 1.12 $\pm$ 0.52 \\
Light cylinder magnetic field, $B_{LC}$ ($10^{5}$ G) & 9.8472 $\pm$ 0.0006 & 2.49 $\pm$ 0.81 \\
\tableline
\end{tabular}
\end{small}
\end{center}
\end{table}

\begin{table}
\begin{center}
\begin{small}
\caption{Light curve and spectral parameters of PSRs B1937+21 and B1957+20 in gamma rays. Details on the measurement of these parameters are given in Section \ref{gamma}. Peak positions $\Phi_i$, full widths at half-maxima FWHM$_i$ and radio-to-gamma-ray lags $\delta_i$ are given in phase units, between 0 and 1. \label{gammaparam}}
\begin{tabular}{lcc}
\tableline\tableline
Parameter & PSR~B1937+21 & PSR~B1957+20 \\
\tableline
First peak position, $\Phi_1$ &  $0.004 \pm 0.009$ & $0.146 \pm 0.026$ \\
First peak full width at half maximum, FWHM$_1$ & $0.030 \pm 0.029$ & $0.137 \pm 0.074$ \\
First peak radio-to-gamma-ray lag, $\delta_1$ & $-0.010 \pm 0.009$ & $-0.016 \pm 0.026$ \\
Second peak position, $\Phi_2$ & $0.543 \pm 0.013$ & $0.616 \pm 0.002$ \\
Second peak full width at half maximum, FWHM$_2$ & $0.041 \pm 0.041$ & $0.014 \pm 0.007$ \\
Second peak radio-to-gamma-ray lag, $\delta_2$ & $0.006 \pm 0.013$ & $0.012 \pm 0.002$ \\
\tableline
Photon index, $\Gamma$ & 1.43 $\pm$ 0.87 $\pm$ 0.40 & 1.33 $\pm$ 0.57 $\pm$ 0.09 \\
Cutoff energy, $E_c$ (GeV) & 1.15 $\pm$ 0.74 $\pm$ 0.43 & 1.30 $\pm$ 0.56 $\pm$ 0.13 \\
Photon flux, $F$ ($\geq 0.5$ GeV) (10$^{-8}$ cm$^{-2}$ s$^{-1}$) & 1.22 $\pm$ 0.23 $\pm$ 0.05 & 0.77 $\pm$ 0.09 $\pm$ 0.01 \\
Energy flux, $G$ ($\geq 0.5$ GeV) (10$^{-11}$ erg cm$^{-2}$ s$^{-1}$) & 1.98 $\pm$ 0.32 $\pm$ 0.04 & 1.34 $\pm$ 0.15 $\pm$ 0.01 \\
Extrapolated photon flux, $F$ ($\geq 0.1$ GeV) (10$^{-8}$ cm$^{-2}$ s$^{-1}$) & 5.97 $\pm$ 4.89 $\pm$ 3.58 & 3.09 $\pm$ 1.62 $\pm$ 0.43 \\
Extrapolated energy flux, $G$ ($\geq 0.1$ GeV) (10$^{-11}$ erg cm$^{-2}$ s$^{-1}$) & 3.63 $\pm$ 1.58 $\pm$ 1.09 & 2.17 $\pm$ 0.54 $\pm$ 0.11 \\
Luminosity, $L_\gamma$ / $f_\Omega$ ($\geq 0.1$ GeV) (10$^{34}$ erg s$^{-1}$) & 25.8 $\pm$ 21.2 $\pm$ 19.6 & 1.62 $\pm$ 1.00 $\pm$ 0.92 \\
Efficiency, $\eta$ / $f_\Omega$ ($\geq 0.1$ GeV) & 0.23 $\pm$ 0.19 $\pm$ 0.18 & 0.22 $\pm$ 0.17 $\pm$ 0.16 \\
\tableline
\end{tabular}
\end{small}
\end{center}
\end{table}

\begin{table}
\begin{scriptsize}
\begin{center}
\caption{Best-fit parameters obtained from the modeling of radio and gamma-ray light curves of PSRs B1937+21 and B1957+20, as discussed in Section \ref{modeling}, for each emission geometry, altitude-limited Two Pole Caustic (alTPC) or altitude-limited Outer Gap (alOG). Altitudes are expressed relative to the light cylinder radius. See the text for more details on the different parameters. \label{fitting_results}}
\begin{tabular}{lcccc}
\tableline\tableline
& \multicolumn{2}{c}{PSR~B1937+21} & \multicolumn{2}{c}{PSR~B1957+20} \\
Parameter & alTPC & alOG & alTPC & alOG \\
\tableline
Magnetic inclination angle, $\alpha$ ($^\circ$) & $75^{+8}_{-6}$ & $84^{+2}_{-6}$ & $47^{+5}_{-13}$ & $31^{+39}_{-3}$ \\

Observer angle, $\zeta$ ($^\circ$) & $80 \pm 3$ & $84^{+1}_{-3}$ & $85^{+1}_{-7}$ & $89^{+5}_{-3}$ \\

Gamma-ray emission gap width, $w_\gamma$ & $0.10 \pm 0.05$ & $0.05 \pm 0.05$ & $0.05 \pm 0.05$ & $0.05 \pm 0.05$ \\

Radio emission gap width, $w_R$ & $0.00 \pm 0.05$ & $0.00 \pm 0.05$ & $0.05 \pm 0.05$ & $0.10 \pm 0.05$ \\

Gamma-ray emission altitudes & $\left[ R_{NS} ; 1\pm0.2 \right]$ & $\left[ R_{NCS} ; 1^{+0.2}_{-0.1} \right]$ & $\left[ R_{NS} ; 1.2^{+0.1}_{-0.4} \right]$ & $\left[ R_{NCS} ; 1.1^{+0.1}_{-0.2} \right]$ \\

Radio emission altitudes & $\left[ 0.7^{+0.1}_{-0.3} ; 0.9^{+0.2}_{-0.1} \right]$ & $\left[ 0.6 \pm 0.1 ; 0.9 \pm 0.1 \right]$ & $\left[ 0.8 \pm 0.1 ; 1.0^{+0.2}_{-0.1} \right]$ & $\left[ 0.7 \pm 0.1 ; 0.9^{+0.2}_{-0.1} \right]$ \\

Geometrical correction factor, $f_\Omega$ & $1.0^{+0.08}_{-0.03}$ & $0.98^{+0.05}_{-0.02}$ & $0.56^{+0.39}_{-0.02}$ & $0.82^{+0.06}_{-0.12}$ \\

Likelihood parameter, $-\ln(L)$ & 126.3 & 130.9 & 123.7 & 128.3 \\
\tableline

Corrected $L_\gamma$ ($\geq 0.1$ GeV) (10$^{34}$ erg s$^{-1}$) & $25.8^{+21.8+20.2}_{-24.8-23.4} $ & $25.2^{+21.0+19.4}_{-27.1-26.0}$ & $2.71^{+1.82+1.70}_{-2.22-2.12}$ & $0.37^{+1.64+1.63}_{-0.35-0.33} $ \\

Corrected $\eta$ ($\geq 0.1$ GeV) & $0.23^{+0.20+0.18}_{-0.23-0.21}$ & $0.23^{+0.19+0.18}_{-0.25-0.24}$ & $0.36^{+0.29+0.28}_{-0.34-0.33}$ & $0.05^{+0.22+0.22}_{-0.05-0.05}$ \\
\tableline
\end{tabular}
\end{center}
\end{scriptsize}
\end{table}%

\end{document}